\documentclass[12pt]{article}
\usepackage{epsfig}
\usepackage{contigrefs}

\begin{document}

\begin{center}
{\bf  \large Going through Rough Times: from Non-Equilibrium Surface Growth to
Algorithmic Scalability}
\end{center}

\vspace*{0.5truecm}
\noindent
G.\ Korniss,$^{1}$
M.A.\ Novotny,$^{2}$
P.A. Rikvold,$^{3}$
H. Guclu,$^1$
and
Z. Toroczkai$^{4}$\\
$^1$Department of Physics, Applied Physics, and Astronomy,
Rensselaer Polytechnic Institute,\\
110 8th Street, Troy, NY 12180-3590 USA\\
$^2$Department of Physics and Astronomy and Engineering Research Center,\\
Mississippi State University, Mississippi State, MS 39762-5167 USA\\
$^3$School of Computational Science and Information Technology,
Department of Physics, and Center for Materials Research and Technology,\\
Florida State University, Tallahassee, FL 32306 USA\\
$^4$Theoretical Division and Center for Nonlinear Studies, MS-B258\\
Los Alamos National Laboratory, Los Alamos, NM 87545 USA

\vspace{0.4truecm}
\noindent {\bf ABSTRACT}

\vspace{0.3truecm}
Efficient and faithful parallel simulation of large asynchronous
systems is a challenging computational problem. It requires using the
concept of local simulated times and a synchronization scheme.
We study the scalability of massively parallel algorithms for
discrete-event simulations which employ conservative
synchronization to enforce causality. We do this by looking at the
simulated time horizon as a complex evolving system, and we identify its
universal characteristics. We find that the time horizon for the
conservative parallel discrete-event simulation scheme exhibits
Kardar-Parisi-Zhang-like kinetic roughening. This implies that the
algorithm is asymptotically scalable in the sense that the average
progress rate of the simulation approaches a non-zero constant. It
also implies, however, that there are diverging memory requirements
associated with such schemes.

\vspace{0.4truecm}
\noindent {\bf INTRODUCTION}

\vspace{0.3truecm}
Faithful and efficient simulation of complex systems with many
interacting degrees of freedom is an important and  challenging
computational task.
In a large class of systems the dynamic evolution is inherently
stochastic and changes in the local configuration of the system occur
randomly in space and time. The modeling of these systems can
follow a ``bottom-up'' approach, starting with the definition of the
``microscopic'' dynamics. Then a master equation can be
constructed with the corresponding transition probabilities.
In most cases, for systems with many interacting degrees of freedom,
even the numerical solution of the master equation (typically involving
the numerical diagonalization of huge matrices) becomes
insurmountable. This is when simulation becomes an invaluable tool for
complex system modeling.

In the physics, chemistry, and biology
communities these types of simulations are most commonly referred to as
dynamic
or kinetic Monte Carlo simulations. In computer science they are
called {discrete-event} simulations. The updates (governed by the
microscopic dynamics) in the (typically local) configuration of the
system are considered discrete events. The basic notion of discrete-event
simulation is that time is continuous and the discrete events occur
instantaneously. Between events, the state (configuration) of the system
remains
unchanged. If the events occur at random instants of
time, the dynamics can be referred to as {\em asynchronous}. Examples
of such systems include the Ising model with the Glauber
or Metropolis dynamics
(the discrete events are the spin-flip attempts), cellular communication
networks (call arrivals), spatial epidemic models (infections),
financial markets (buy/sell orders), or internet traffic (packet
transmission/reception).  In this paper we focus on systems
with general but {\em short-ranged} interactions on
{\em regular lattices} and assume that the event dynamics can be described
as a superposition of a large number of independent
Poisson processes running in parallel ({\em Poisson asynchrony}).

When the size of the system becomes large, parallelization may be
needed to obtain results within an acceptable time frame. Massively
parallel simulation for complex systems with asynchronous dynamics,
i.e., {\em parallel discrete-event simulations} (PDES), is a standard
technique among computer scientists. It is somewhat
surprising that despite PDES having a long history as far
as applications and scalability are concerned \cite{FUJI90,NICOL94},
very few of the PDES
techniques have filtered through, e.g., to the physics community. Even
the simplest random-site update Monte Carlo schemes \cite{BIND95}
where update attempts converge to Poisson arrivals in the large
system-size limit, were long believed to be inherently serial.
In this regard, Lubachevsky's work \cite{LUBA87,LUBA88}
was rather illuminating, by
illustrating how to apply the PDES scheme to the Ising
model on a regular lattice with Glauber dynamics \cite{BIND95,GLAUB63}.

In a PDES scheme each processing element (PE) carries a subsystem of
the full system via simple spatial decomposition.
The difficulty of PDES is that the discrete events are not
synchronized by a global clock since the dynamic is asynchronous. To
put it simply, the paradoxical task is to (algorithmically)
parallelize (physically) non-parallel
dynamics. To achieve this, one must use the concept of local simulated
times (or virtual times) and a synchronization scheme.
The parallel algorithm must
concurrently advance the local simulated time on each subsystem
carried by a PE, without violating causality.
In a ``conservative'' PDES scheme \cite{CHAND79,CHAND81},
only those PEs that are guaranteed
not to violate causality are allowed to process their events and
increment their local time. The rest of the PEs must ``idle.''
In an ``optimistic'' approach \cite{JEFF85},
the PEs do not have to idle, but since
causality is not guaranteed at every update, the simulated history on
certain PEs can become corrupted. This requires a complex ``rollback''
protocol to correct erroneous computations. Both simulation approaches
lead to an evolving and fluctuating time horizon during algorithmic
execution.

For systems which can be modeled on regular lattices with short-ranged
interactions, the conservative scheme can be highly
efficient. Recently, it was implemented for modeling magnetization
switching \cite{KORN99_JCP} and the dynamic phase transition
\cite{KORN01_PRE} in highly anisotropic thin-film ferromagnets.
For example, the nearest-neighbor interaction implies that in
order to ensure causality, PEs need to exchange their local simulated
(or virtual) times only with ``neighboring'' PEs in the virtual PE topology.
Communication times between PEs and possible idling due to the
conservative synchronization protocol can be greatly suppressed by each
PE carrying a large block of sites (spins)
\cite{LUBA87,LUBA88,KORN99_JCP}, yielding encouraging efficiencies and
utilizations (fraction of non-idling PEs).
Since there is a finite number of PEs in any architecture, one can
typically be satisfied with these ``experimental'' observations.
However, as the number of PEs available to simulating complex systems
increases to tens of thousands, scalability questions become fundamental.
Further, questions, such as how the utilization behaves in the
asymptotic limit when the number of PEs goes to infinity, truly lie
at the heart of any PDES scheme.

In this paper we address fundamental scalability
questions for the general conservative parallel simulations for systems
on regular lattices with short-range interactions.
The way we tackle the problem, in some sense, goes opposite to the
usual flow of a typical scientific modeling and simulation process. There,
one develops advanced and sophisticated computational algorithms
to study and understand systems in Nature. Here, by
knowing how certain natural systems behave, we try to understand how
advanced algorithms, PDES in particular, work: whether or not they
are scalable, and how they can be optimized. Our approach is facilitated by
a mapping \cite{KORN00_PRL,KORN00_UGA} between non-equilibrium surface
growth and the progress of the simulation (the evolution of the
simulated time horizon). At the end, of course, based on the knowledge
gained
after answering fundamental scalability questions, one hopes to close this
``loop'' by devising and optimizing PDES schemes that can be used to
investigate challenging problems in natural, artificial, or social systems.
Analogies of a similar kind, e.g., that between phase transitions and
computational complexity \cite {COMPLEX,KIRK99} also have
turned out be highly fruitful to gain more
insight in traditionally difficult problems in computer science.
Also, exploiting analogies
between the evolution of the time horizon and that of known {\em
physical} systems appears to be rather helpful in understanding the
performance for optimistic schemes as well.
There is some evidence \cite{OVER00,OVER01b} that the
time horizon in rollback-based schemes can exhibit self-organized
criticality and power-law spatio-temporal correlations, which can
be crucial to extract the scalability properties.

Based on our mapping between non-equilibrium surface growth and
the progress of the simulation, it is clear that the tools and frameworks of
modern statistical physics, in particular those of non-equilibrium
interface/surface growth \cite{BARA95,ZHANG95,KRUG97},
can be extremely helpful in analyzing and
understanding the asymptotic scalability properties of PDES schemes.
To this end, one must look at the
simulation scheme itself as an evolving interacting system of individual PEs
where the synchronization rules among the PEs constitute the effective
interaction.
The evolution of this simulated time horizon, in particular its
average progression rate and statistical spread, will determine the
scalability properties of the corresponding PDES scheme.

In the next section we give an overview
of the basic conservative PDES scheme \cite{LUBA87,LUBA88}. Then we
characterize the morphological properties of the evolving random surface
associated with the simulated time horizon. These findings
yield direct implications for the scalability of the conservative
algorithm for PDES.

\vspace{0.4truecm}
\noindent {\bf THE BASIC CONSERVATIVE APPROACH}

\vspace{0.3truecm}
We consider a $d$-dimensional hypercubic regular lattice topology,
where the underlying physical system has only nearest-neighbor
interactions. However, our results hold for any short-range regular 
interaction pattern. In this paper we consider the case of simple
Poisson asynchrony. Update attempts at each site are identical and
independent
Poisson processes (thus, the random simulated time increments between
two successive update attempts are exponentially distributed) and are
also {\em independent of the state of the underlying physical system}.
The consequence of the former is that the evolution of the simulated
time horizon completely {\em decouples} from the behavior and evolution of
the underlying physical system. For simplicity, we discuss in detail the
``worst-case'' scenario, in which each PE carries one site
(e.g., one spin). In this basic conservative scheme, each PE generates
its own local simulated time for the next update attempt. (The actual
update probabilities depend on the underlying systems, e.g., through
energetics, but they do not affect the evolution of the time horizon.)

The set of local simulated times for $N$ PEs,
$\{\tau_i(t)\}_{i=1}^{N}$,
constitutes the simulated time horizon. Here $t$ is the discrete
number of parallel steps simultaneously performed on each PE
directly related to real/wall-clock time, or if the architecture
operates in an asynchronous execution mode, $t$ is simply the
continuous real time. On a regular $d$-dimensional hypercubic
lattice $N$$=$$L^d$, where $L$ is the linear size of the lattice.
In physics applications one typically specifies the initial
configuration (i.e., at $\tau$=0) of the underlying physical
system. This translates to $\tau_i(0)$$=$$0$ for every  site for the
initial condition of the parallel simulation. Then at each
parallel update, {\em only} those PEs for which the local
simulated time is {\em not greater} than the local simulated times
of their nearest neighbors, can increment their local time by an
exponentially distributed random amount, $\eta_i(t)$. The other PEs
must idle. Without loss
of generality we take independent, identically distributed (iid)
exponential variables of mean one
in simulated time units (stu), $\langle\eta_i(t)\rangle$$=$$1$.
Due to the continuous nature of the random simulated times, for $t>0$ the
probability of equal-time updates for any two sites is of measure
zero. The comparison with nearest neighbor simulated times and, if
necessary,
idling enforces causality. Also, at worst, the PE
with the global minimum simulated time can make progress, so the
algorithm is free from deadlock. For this basic conservative
scheme, the theoretical efficiency or utilization (ignoring
communication overheads) is simply the (average) fraction of non-idling PEs.
This corresponds to the density of local minima of the simulated
stochastic time horizon, which determines the average progress rate
of the simulation. It can be written as
\begin{equation}
\left\langle u(t) \right\rangle_N = \left\langle \sum_{i=1}^{N}
\prod_{j\in D^{\rm nn}_{i}}\!\Theta\left(\tau_{j}(t)-\tau_{i}(t)\right)
\right\rangle \;,
\label{util_def}
\end{equation}
where $D^{\rm nn}_{i}$ is the set of nearest neighbors of
site $i$, $\Theta(\cdot)$ is the Heaviside step function, and the
$\langle\ldots\rangle$ denotes an ensemble average, i.e., an average
over many independent simulations.

Another important aspect of the simulation is
the width of the distribution of the local simulated times.
This property can have serious effects on the ``measurement part''
of the algorithm, e.g. when one attempts to collect and compute
simple statistics for the full underlying physical system ``on the
fly.'' Therefore, one must determine the {\em statistical spread}
(width) of the time horizon as was pointed out in Ref.
\cite{GREEN96}. This quantity can be characterized by
\begin{equation}
\left\langle w^{2}(t) \right\rangle_{N} =
\left\langle \frac{1}{N}\sum_{i=1}^{N}
\left[ \tau_i(t)-\bar{\tau}(t) \right]^2 \right\rangle \;,
\label{width_def}
\end{equation}
where $\bar{\tau}(t)$$=$$(1/N)\sum_{i=1}^{N} \tau_i(t)$ is the mean progress
(``height'') of the time horizon.
The behavior of the width, $\langle w^{2}(t) \rangle$, 
alone typically captures and identifies the
universality class of the non-equilibrium growth process
\cite{BARA95,ZHANG95,KRUG97}.

\vspace{0.4truecm}
\noindent {\bf EVOLUTION OF THE SIMULATED TIME HORIZON}

\begin{figure}[t]
\vspace*{3.0truecm}
\includegraphics{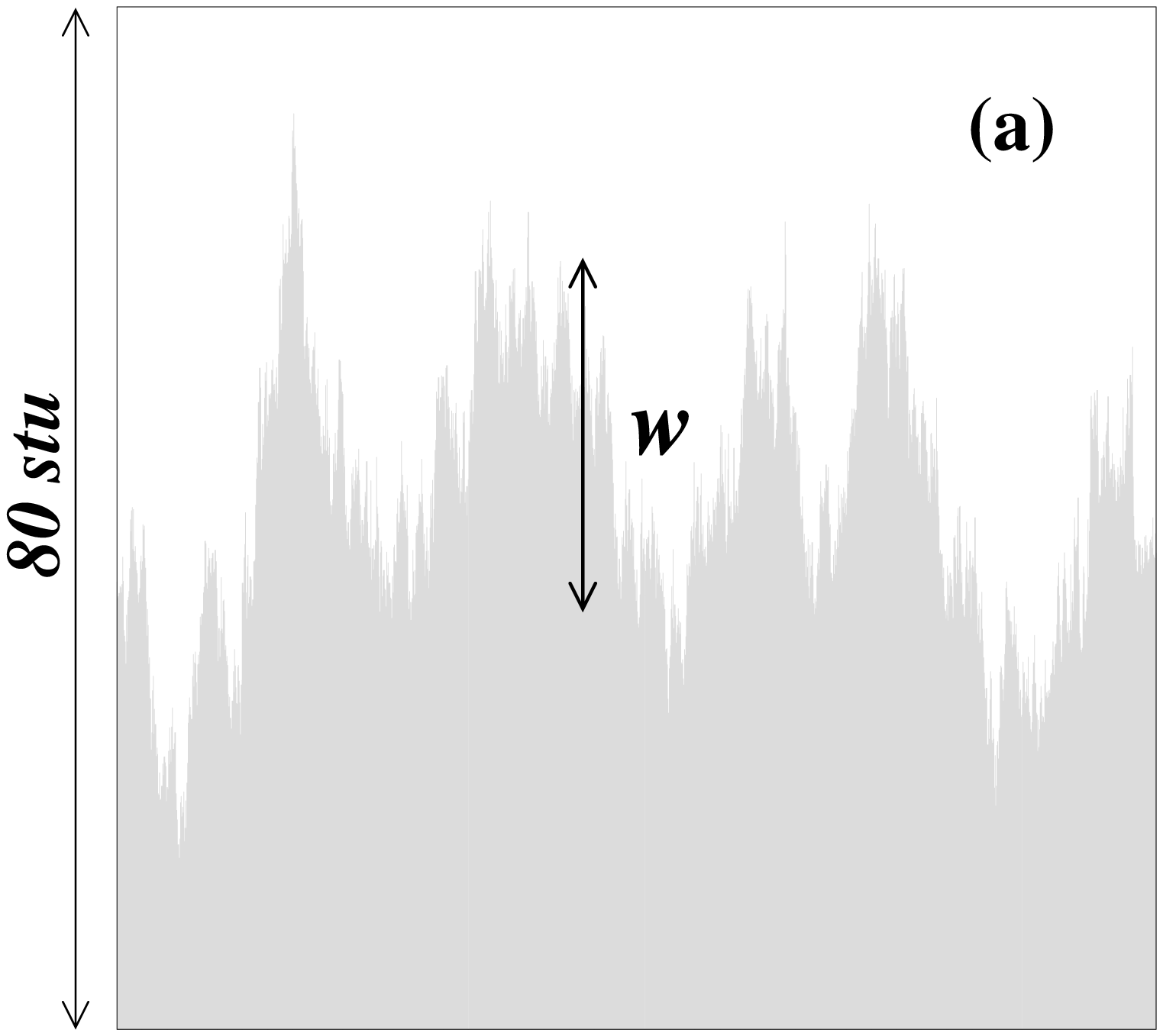}
\includegraphics{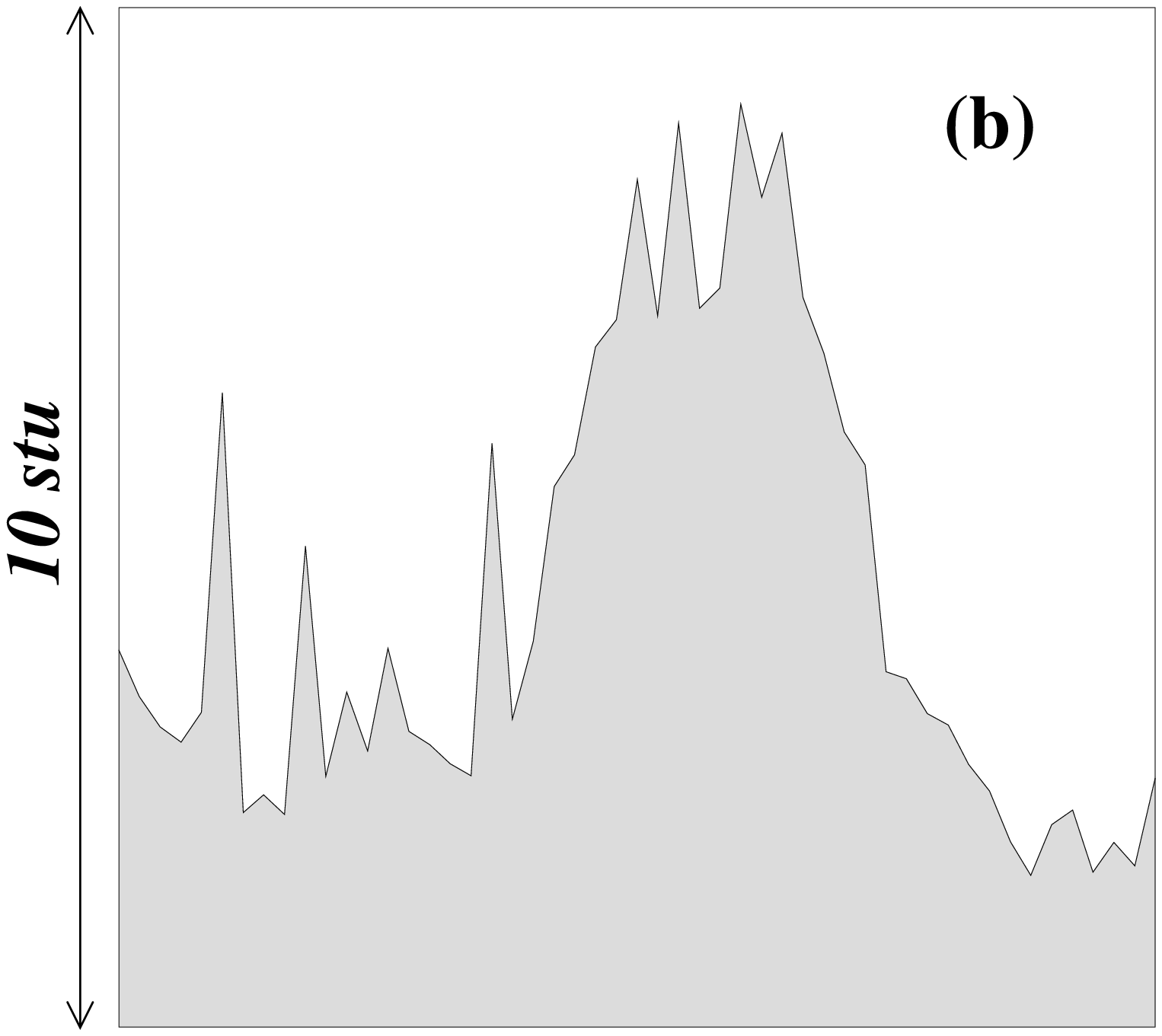}
\vspace*{1.5truecm}
\caption{\small
(a) Snapshot configuration of the actual simulated time
horizon for the one-dimensional one site per PE topology with
$L$$=$$10,000$ PEs. The time horizon propagates ``upwards'' in the figure,
and $w$ indicates the average width of the time horizon.
(b) A short segment ($50$ sites) of the same configuration. The arrows on
the left indicate the vertical scales in simulated time units (stu).}
\label{fig1}
\end{figure}

\begin{figure}[t]
\vspace*{3.0truecm}
\includegraphics{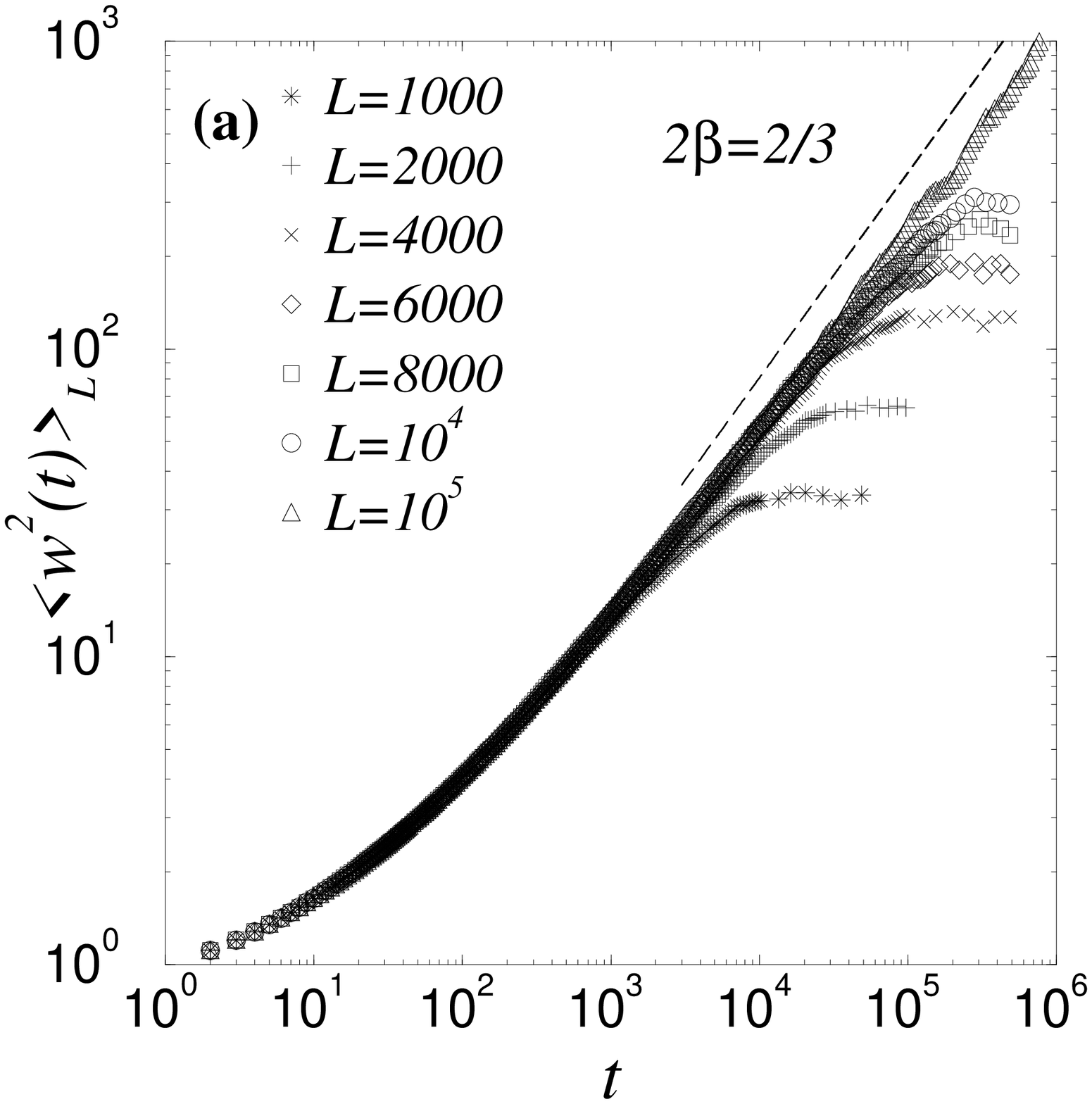}
\includegraphics{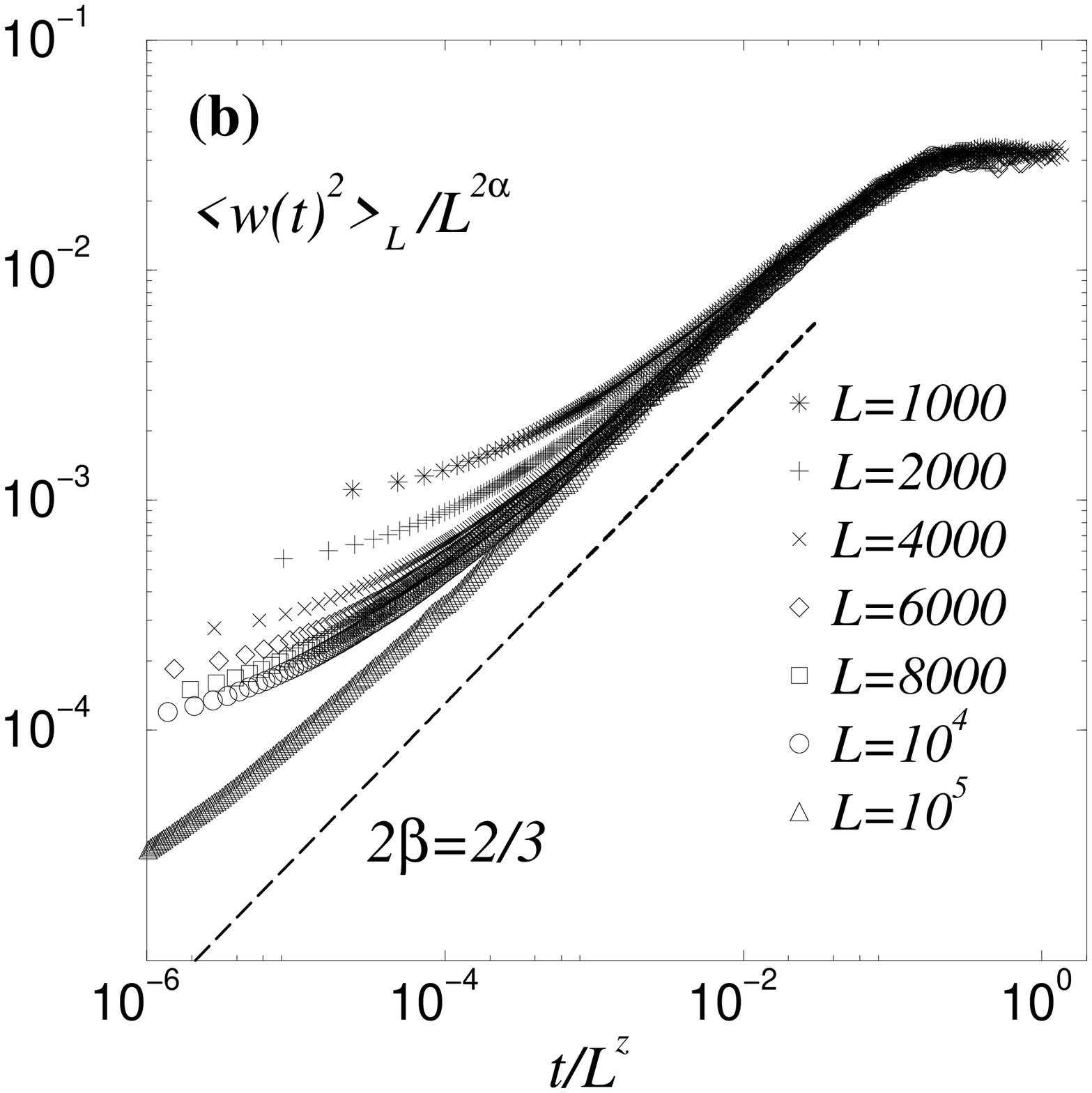}
\includegraphics{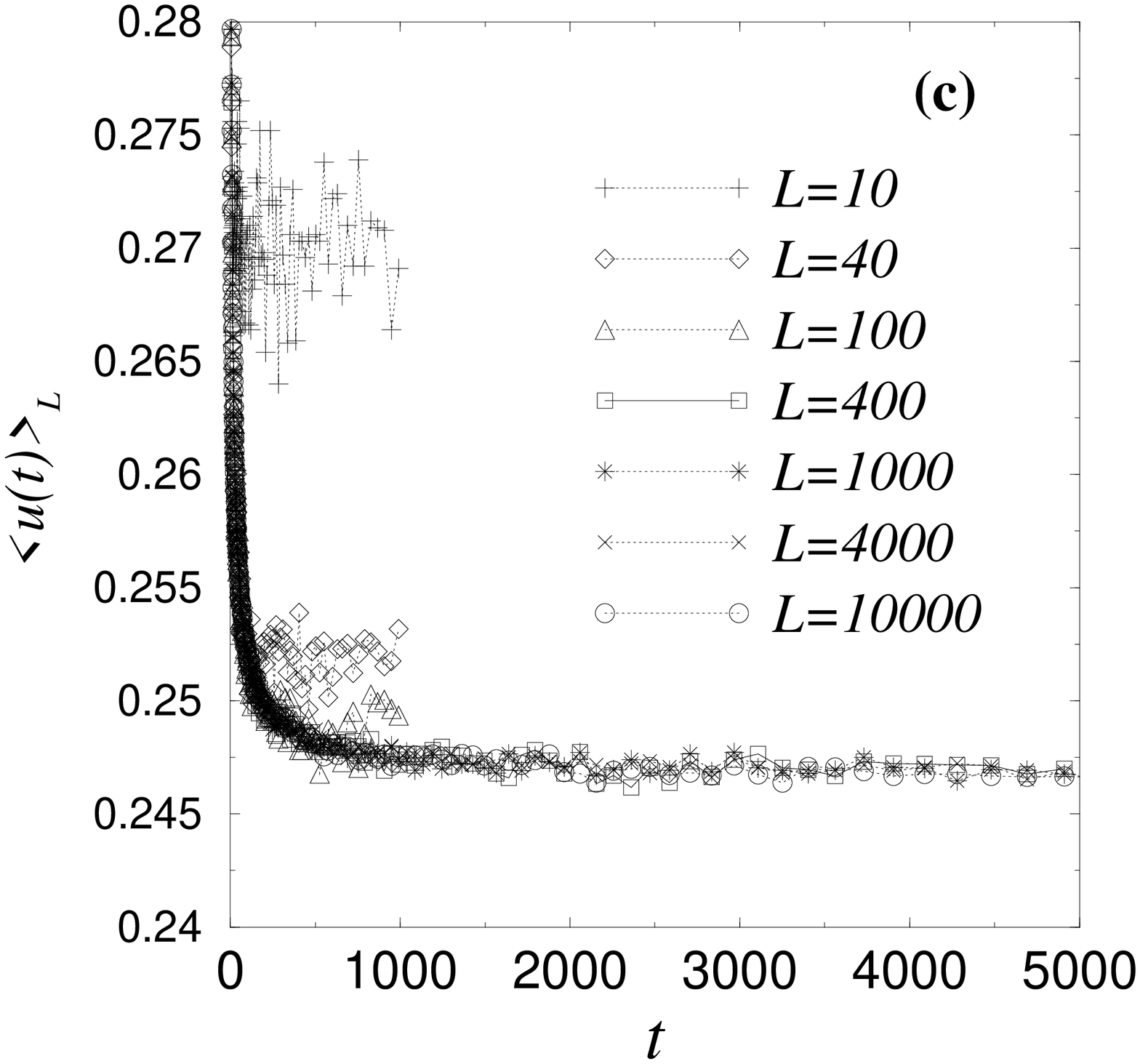}
\vspace*{2.5truecm}
\caption{\small
(a) Kinetic roughening of the simulated time horizon for the
one-dimensional one site per PE basic conservative scheme.
Note the log-log scales, indicating the power-law growth of the width before
saturation. The dashed line corresponds to a power law with the exact KPZ
exponent $2\beta$$=$$2/3$.
(b) The same behavior as in (a), using rescaled variables to demonstrate
the dynamic scaling hypothesis, Eq.~(\protect\ref{Vicsek}).
(c) Time dependent utilization for various system sizes.}
\label{fig2}
\end{figure}

\vspace{0.3truecm}
The conservative synchronization protocol together with the virtual
communication topology of the PEs (mimicking the interaction topology
of the underlying physical system) fully specify the
``microscopic'' dynamics of the growth process associated with the
evolution of the time horizon. The rules of the conservative
synchronization can be compactly summarized as
\begin{equation}
\tau_{i}(t+1) =   \tau_{i}(t)
 + \prod_{j\in D^{\rm nn}_{i}}\!\Theta\left(\tau_{j}(t)-\tau_{i}(t)\right)
\eta_{i}(t) \;.
\label{tau_evol}
\end{equation}
The above equation simply reflects that the local simulated time at site
(PE)
$i$ is incremented by an exponentially distributed random amount
$\eta_i(t)$,
{\em provided} that site $i$ is a local minimum of the time horizon.
This stochastic time evolution equation is exact for the basic conservative
scheme and can be easily simulated on a serial (!)
workstation. Snapshots of the evolving and
fluctuating time horizon obtained from direct simulations
are shown in Fig.~\ref{fig1}.
We discuss in detail the one-dimensional case ($N$$=$$L$) with
periodic boundary conditions (i.e., ring topology).
Replacing the Heaviside step function with a limiting smooth representation,
one can perform standard coarse graining on Eq.~(\ref{tau_evol})
\cite{KORN00_PRL}, yielding
\begin{equation}
\frac{\partial\hat{\tau}}{\partial t} =
\frac{\partial^2 \hat{\tau}}{\partial x^2} -
\lambda \left(\frac{\partial \hat{\tau}}{\partial x}\right)^2
+ \hat{\eta}(x,t) \;.
\label{KPZ_eq}
\end{equation}
Here $\hat{\tau}$ is the coarse-grained ``height'' fluctuation,
and the temporal and spatial derivatives are just the naive continuum
interpretations of the differences, e.g.,
$\partial\hat{\tau}/\partial x$$=$$\tau_i - \tau_{i-1}$. Similarly,
$\hat{\eta}$
is the coarse-grained noise, delta correlated in space and time.
Higher order terms in Eq.~(\ref{KPZ_eq}) are dropped since they are
irrelevant
in a Renormalization Group (RG) sense. Equation ~(\ref{KPZ_eq}) is the
Kardar-Parisi-Zhang (KPZ) equation \cite{KARD86},
which has turned out to be of central importance and have a wealth of
applications for numerous artificial and natural growth processes over the
past
two decades, including molecular beam epitaxy, electrochemical deposition,
fluid
flow in porous media, and growth of bacterial colonies
\cite{BARA95,ZHANG95}.
Thus, we expect that the simulated time horizon exhibits
{\em kinetic roughening}, the main feature of KPZ growth.
Indeed, our direct simulation for the time horizon evolution,
Eq.~(\ref{tau_evol}), confirms the coarse-graining approach
[Fig.~\ref{fig2}(a) and~(b)].
There is a system-size dependent characteristic time scale, the crossover
time,
$t_\times$$\sim$$L^z$. For very early times, mostly microscopic
details of the dynamics influence the width. For intermediate
times, while $t$$\ll$$t_\times$ still, the width grows
as a power law $\langle w^2(t)\rangle_L$$\sim$$t^{2\beta}$, where
$\beta$ is the {\em growth exponent}. For late times,
$t$$\gg$$t_\times$, the width {\em saturates} for any finite
system size. In this regime the surface reaches a steady-state
evolution, and the fluctuations about the mean are stationary. The
saturation or steady-state value of the width, however, scales as
a power law with the system size, $\left\langle
w^2(\infty)\right\rangle_L$$\sim$$L^{2\alpha}$, where $\alpha$ is the
{\em roughness exponent}.
The time horizon through its progress exhibits exactly the above
scaling behavior with $\beta$$\simeq$$1/3$ and $\alpha$$\simeq$$1/2$,
consistent with the exact one-dimensional KPZ exponents
\cite{BARA95,ZHANG95,KARD86} [Fig.~\ref{fig2}(a)]. This type of temporal and
system-size scaling is consistent with the dynamic scaling
hypothesis \cite{BARA95,VICSEK85} and can be expressed through the
Family-Vicsek scaling relation
\begin{equation}
\left\langle w^2(t)\right\rangle_L = L^{2\alpha}f(t/L^{z})
\label{Vicsek}
\end{equation}
together with the important scaling law, $\alpha$$=$$\beta z$.
Note that the scaling function $f(x)$ depends on $t$ and the linear system
size $L$ only through the specific combination $t/L^z$, reflecting the
importance of the crossover time $t_\times$. For small
values of its argument $f(x)$ behaves as a power law, while for large
arguments it approaches a constant
\begin{equation}
f(x) \sim
\left\{
\begin{array}{ll}
x^{2\beta}    & \mbox{if $x$$\ll$$1$} \\
\mbox{const.} & \mbox{if $x$$\gg$$1$}
\end{array}
\right. \;.
\label{f_scaling}
\end{equation}
The existence of the above scaling function implies that if one
plots the rescaled variables $\langle w^2(t)\rangle_L/L^{2\alpha}$
vs $t/L^z$, then curves for different system sizes collapse for
intermediate and late times. We have confirmed this data collapse for
the simulated time horizon [Fig.~\ref{fig2}(b)].

Direct simulation results for the average rate of progress of the time
horizon
(equivalent to the
utilization in the PDES algorithm) for various system sizes are shown in
Fig.~\ref{fig2}(c). The utilization
$\langle u(t)\rangle_L$ decreases monotonically with time towards a
long-time asymptotic limit well separated from zero,
$\langle u(\infty) \rangle_{\infty}$$\approx$$0.25$.
The fact that it cannot vanish in the infinite system-size limit
can be argued based on an important universal
feature of KPZ-like surfaces:
The steady-state KPZ surface in one dimension is governed by the
Edwards-Wilkinson Hamiltonian \cite{EW82}, i.e., it is
essentially a random-walk profile \cite{BARA95}.
At coarse-grained length scales the local {\em slopes} become
{\em independent}, yielding a {\em non-zero} average density of local
minima, i.e., a non-zero average rate of progress of the
simulation in the $L$$\to$$\infty$ limit in the steady-state.

\begin{figure}[t]
\vspace*{3.0truecm}
\includegraphics{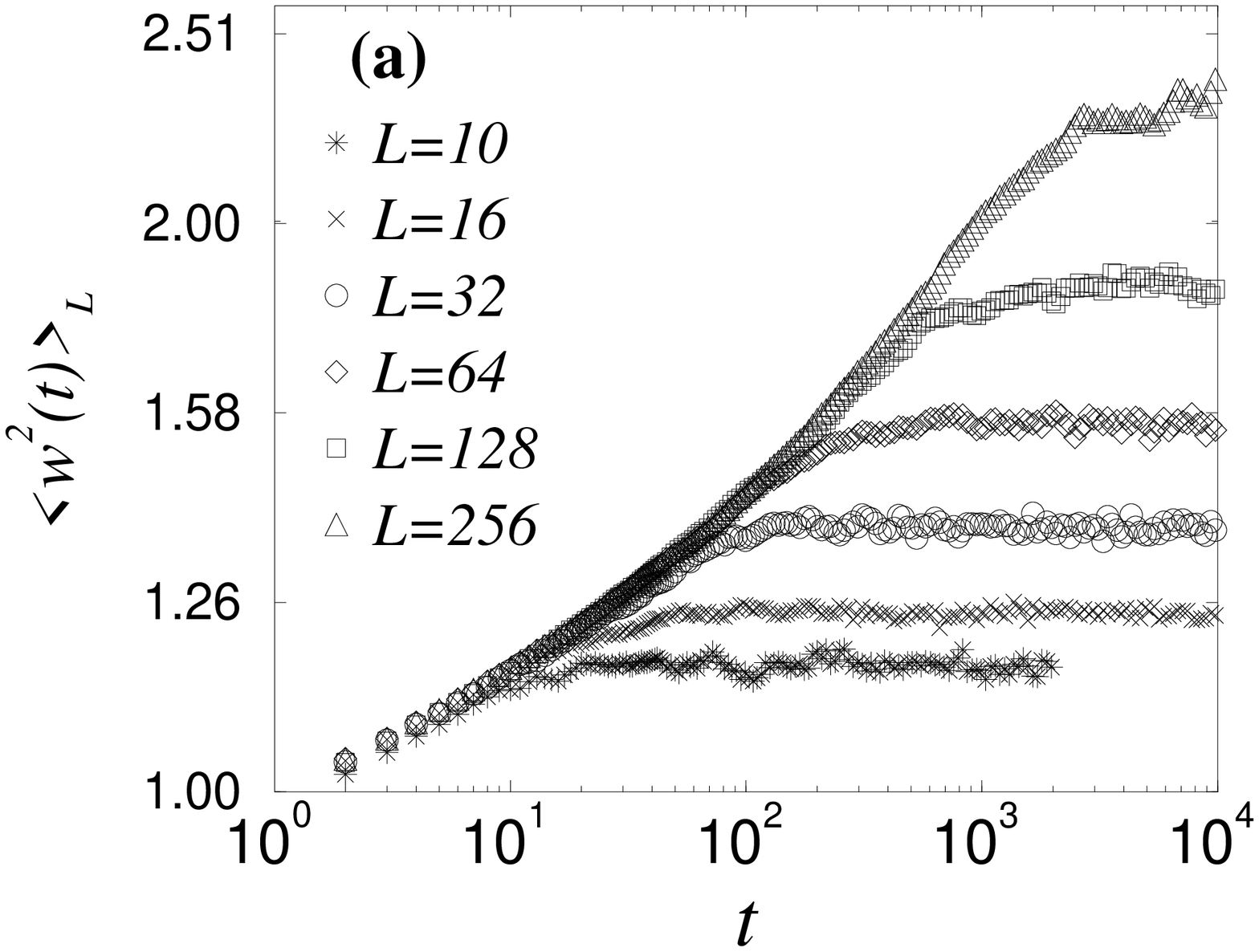}
\includegraphics{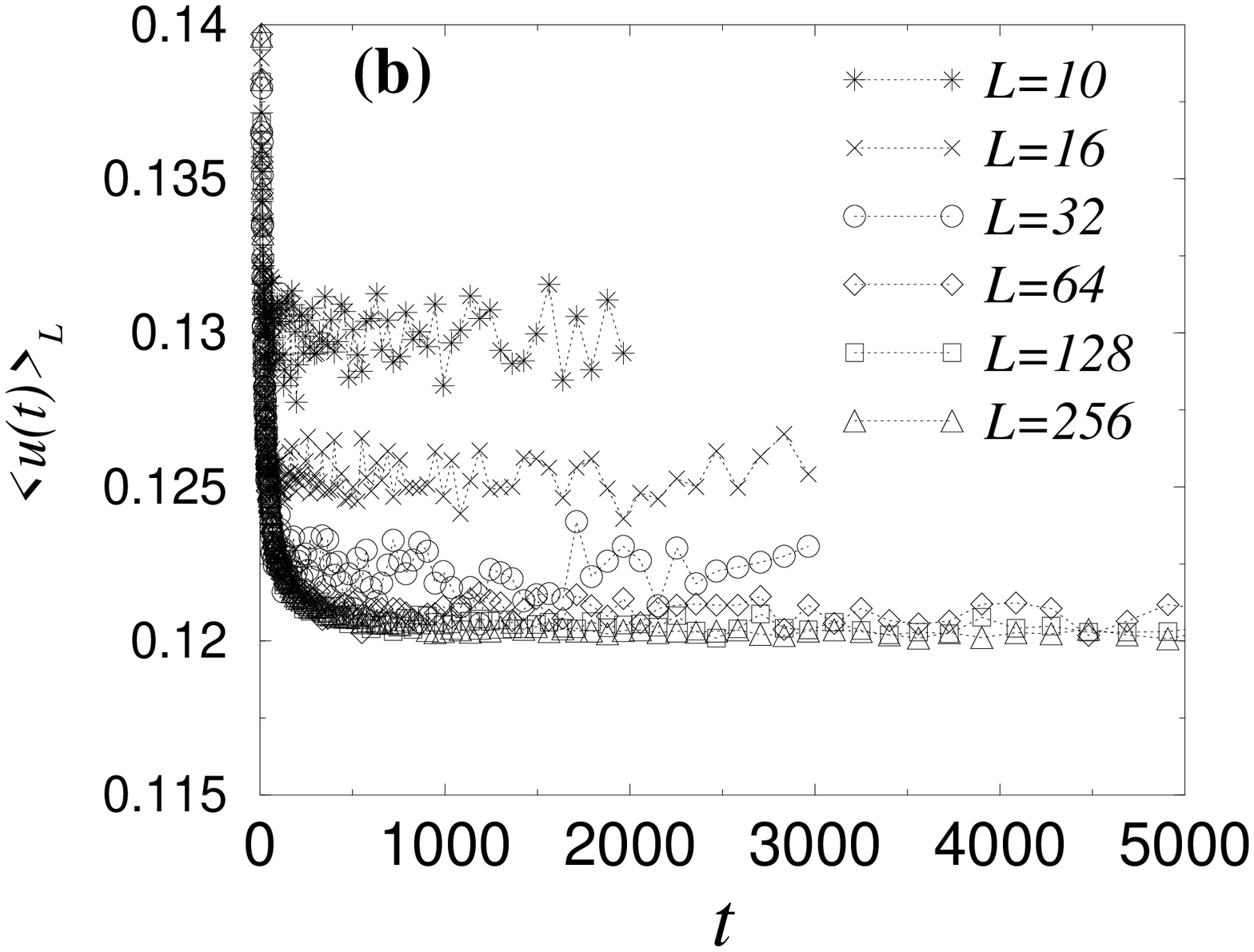}
\includegraphics{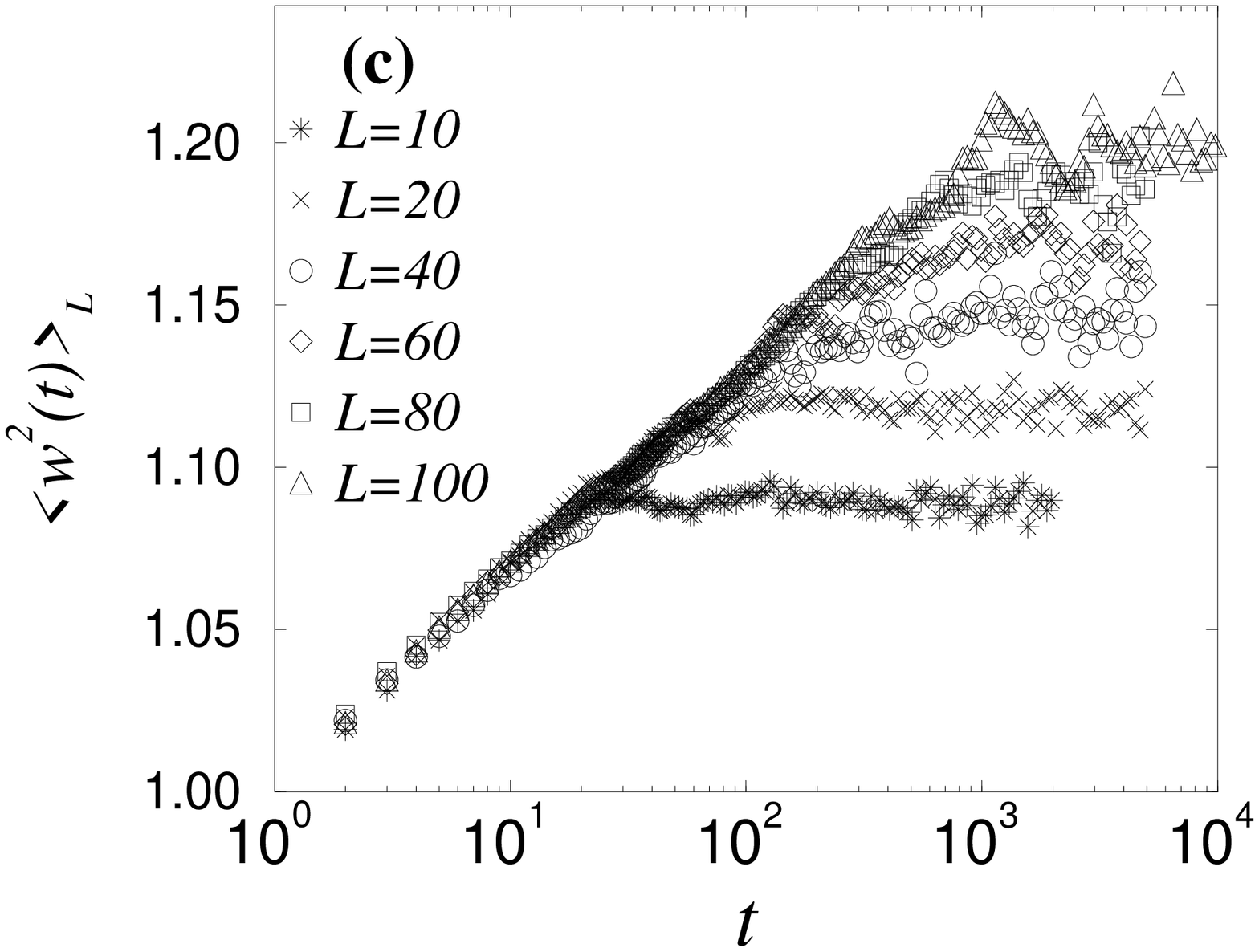}
\includegraphics{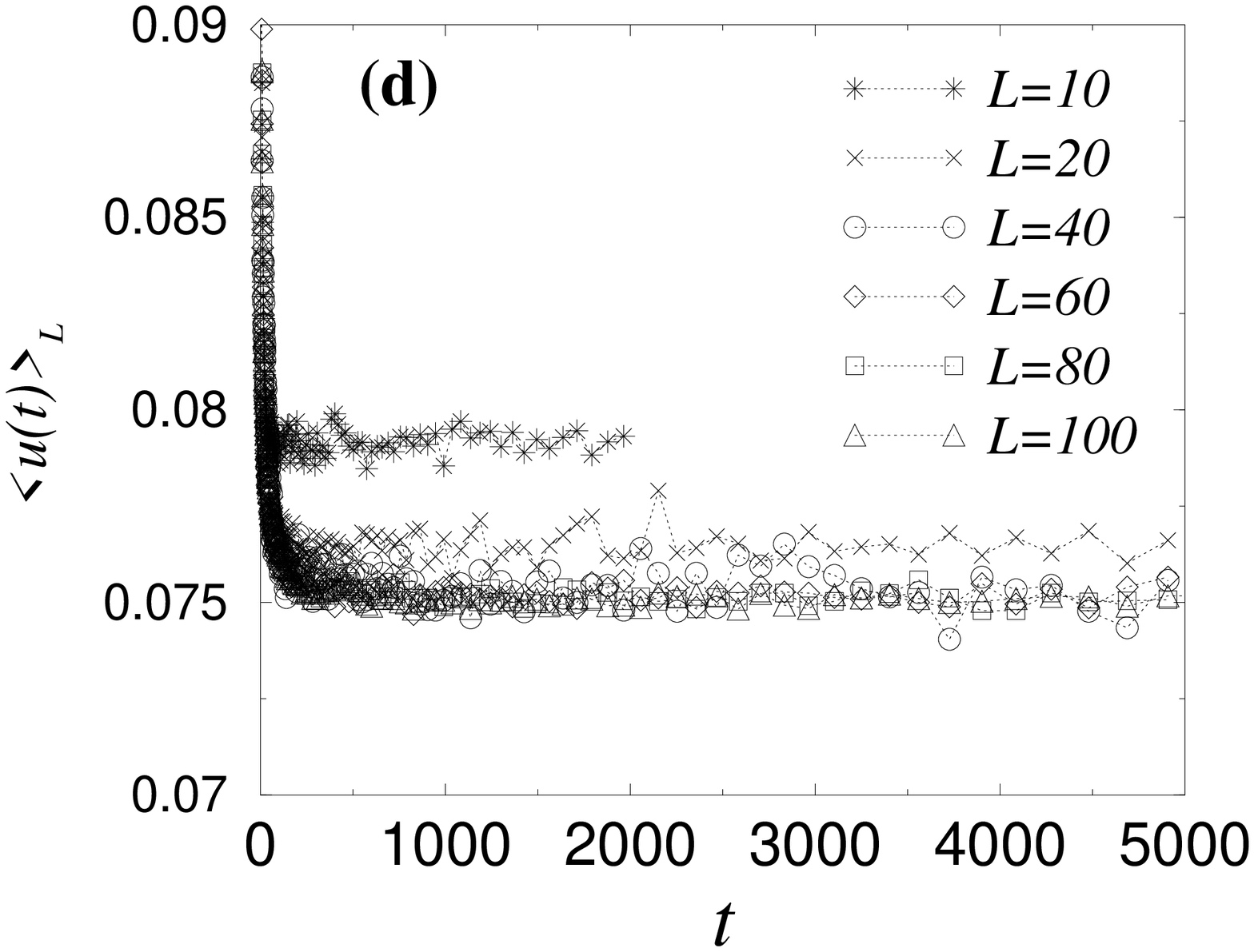}
\vspace*{7.50truecm}
\caption{\small
Evolution of the simulated time horizon in $d$$=$$2$ and $d$$=$$3$:
(a) surface width in $d$$=$$2$; (b) density of local minima in $d$$=$$2$;
(c) surface width in $d$$=$$3$; (d) density of local minima in
$d$$=$$3$.}
\label{fig3}
\end{figure}

In higher dimensions we observe the same qualitative behavior as for
$d$$=$$1$ \cite{KORN00_UGA}.
The surface roughens and {\em saturates} for any finite
system, as seen in Fig.~\ref{fig3}(a) and~(c).
Simultaneously, the density of local minima  decreases monotonically
towards its asymptotic ($t$$\rightarrow$$\infty$) finite-size value
[Fig.~2(b) and~(d)]. Again, the steady-state density of local minima appears
to be well separated from zero. For $d$$=$$2$
$\langle u(\infty) \rangle_{\infty}$$\approx$$0.12$,
and for $d$$=$$3$
$\langle u(\infty) \rangle_{\infty}$$\approx$$0.075$. The
$\langle u(\infty) \rangle_{\infty}$$\sim$${\cal O}(1/K)$
behavior appears to be rather general \cite{GREEN96},
where $K$$=$$2d$ is the number of nearest neighbors on a regular lattice.
Similar to the $d$$=$$1$ case, corrections to scaling are very strong,
both for the surface width and the density of local minima. While for
$d$$=$$1$
we were able to simulate large systems ($L$$\gg$$10^3$) to obtain the KPZ
scaling exponents and the steady-state finite-size behavior of
$\langle u(\infty) \rangle_L$,
in higher dimensions the relatively small system sizes
prevented us from extracting the scaling behavior of the width and the
finite-size effects of the density of local minima.
We conjecture that the simulated time horizon exhibits
KPZ-like evolution in higher dimensions as well.

Next, we investigate in detail the steady-state scaling properties
of the simulated time horizon for $d$$=$$1$, which directly translates to the
asymptotic scalability properties of the corresponding PDES.

\vspace{0.4truecm}
\noindent {\bf SCALING AND SCALABILITY}

\begin{figure}[t]
\vspace*{3.0truecm}
\includegraphics{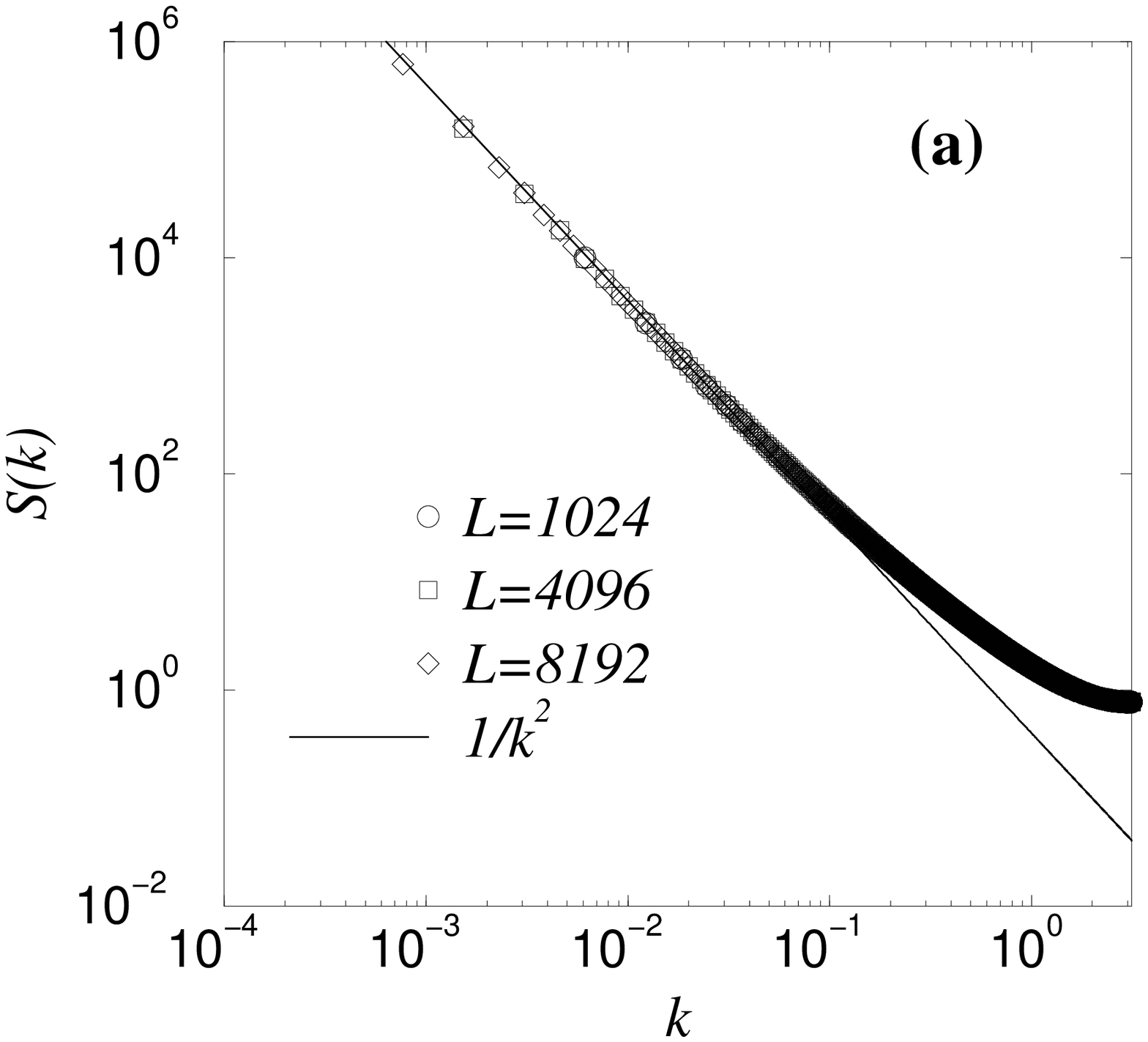}
\includegraphics{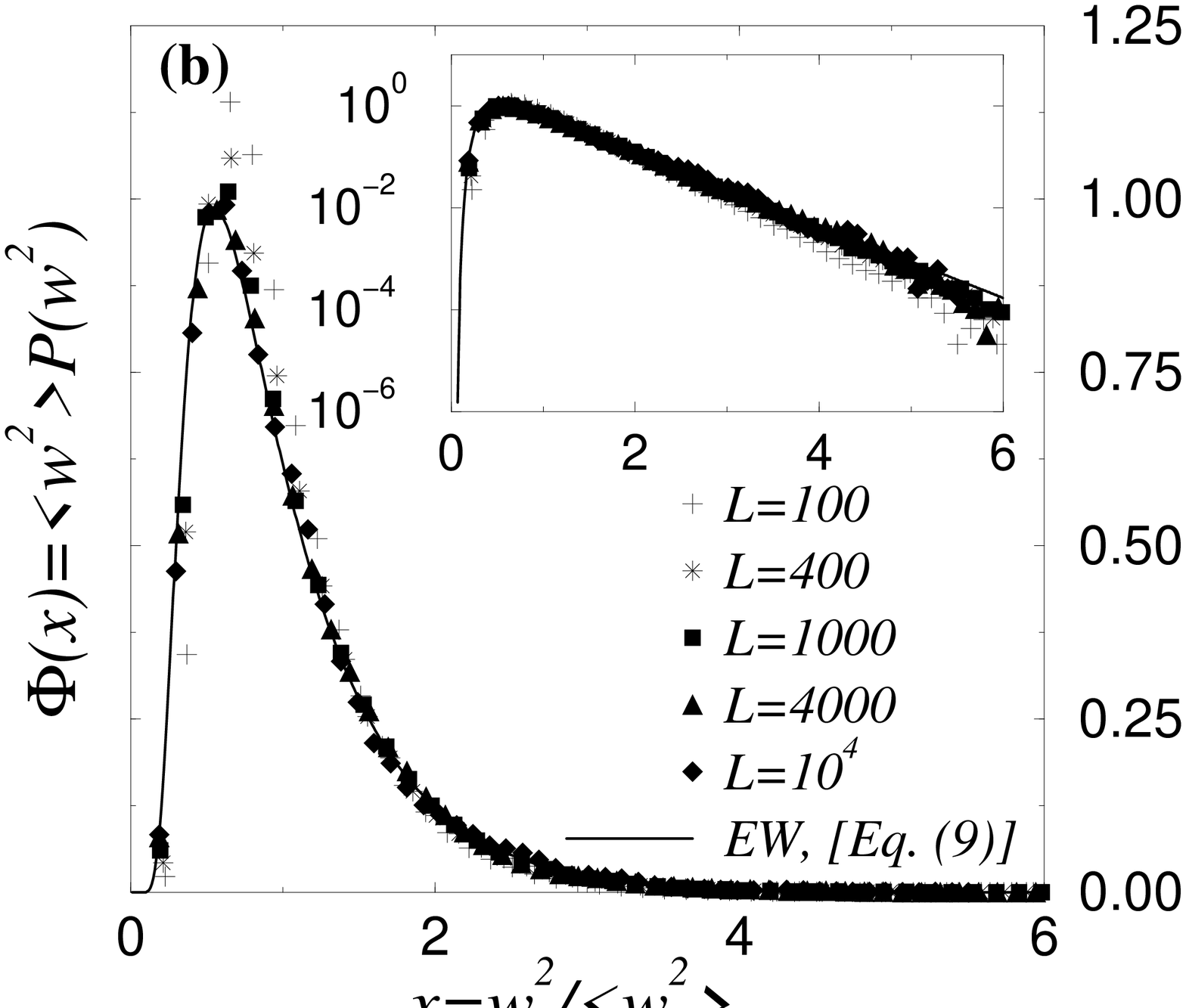}
\vspace*{3.2truecm}
\caption{\small
(a) Steady-state structure factor of the simulated time
horizon. The straight solid line corresponds to the theoretical
prediction Eq.~(\protect\ref{str}).
(b) Steady-state width distribution (inset: on a linear-log scale). The
solid curve is the universal scaling function for the EW class,
Eq.~(\protect\ref{exact_hist}).
Both graphs
are for the one-dimensional one site per PE basic conservative scheme.}
\label{fig4}
\end{figure}

\vspace{0.3truecm}
First, we provide further numerical evidence that
the time horizon belongs to the KPZ universality class,
in particular, in one dimension in the steady state,
it is governed by the Edwards-Wilkinson (EW) Hamiltonian \cite{EW82},
\begin{equation}
{\cal H}_{\rm EW} \propto \frac{1}{2}\int\!\!dx
(\partial \hat{\tau}/\partial x)^2 .
\label{EW}
\end{equation}
Then for small wave-vectors $k$, the average steady-state structure
factor of the surface should behave as
\begin{equation}
S(k)=\langle \tau_k \tau_{-k}\rangle/L \propto 1/k^2\;,
\label{str}
\end{equation}
where $\tau_k$$=$$\sum_{j=1}^{L}e^{-ikj}\tau_{j}$
is the spatial Fourier transform of the time horizon.
This expectation is confirmed by direct simulations as shown in
Fig.~\ref{fig4}(a).

To further probe the universal properties of the surface in the steady
state, we construct the width distribution $P(w^2)$, i.e.,
not just the average but the {\em full} probability density of the
quantity in the brackets in Eq.~(\ref{width_def}) \cite{KORN00_PRL}.
This is a very powerful universal signature of rough surfaces, and in one
dimension it can be tested against analytic results.
The EW class is characterized by a universal scaling function, $\Phi(x)$,
such
that $P(w^2)=\langle w^2\rangle^{-1}\Phi(w^2/\langle w^2\rangle)$
\cite{FORWZ94}, where
\begin{equation}
\Phi(x)=\frac{\pi^2}{3}\sum_{n=1}^{\infty}(-1)^{n-1} n^{2}
e^{ -\frac{\pi^2}{6} n^{2} x } \;.
\label{exact_hist}
\end{equation}
Systems with $L$$\geq$$10^3$ show convincing data collapse
[Fig.~\ref{fig4}(b)] onto this exact scaling function.

\begin{figure}[t]
\vspace*{3.0truecm}
\includegraphics{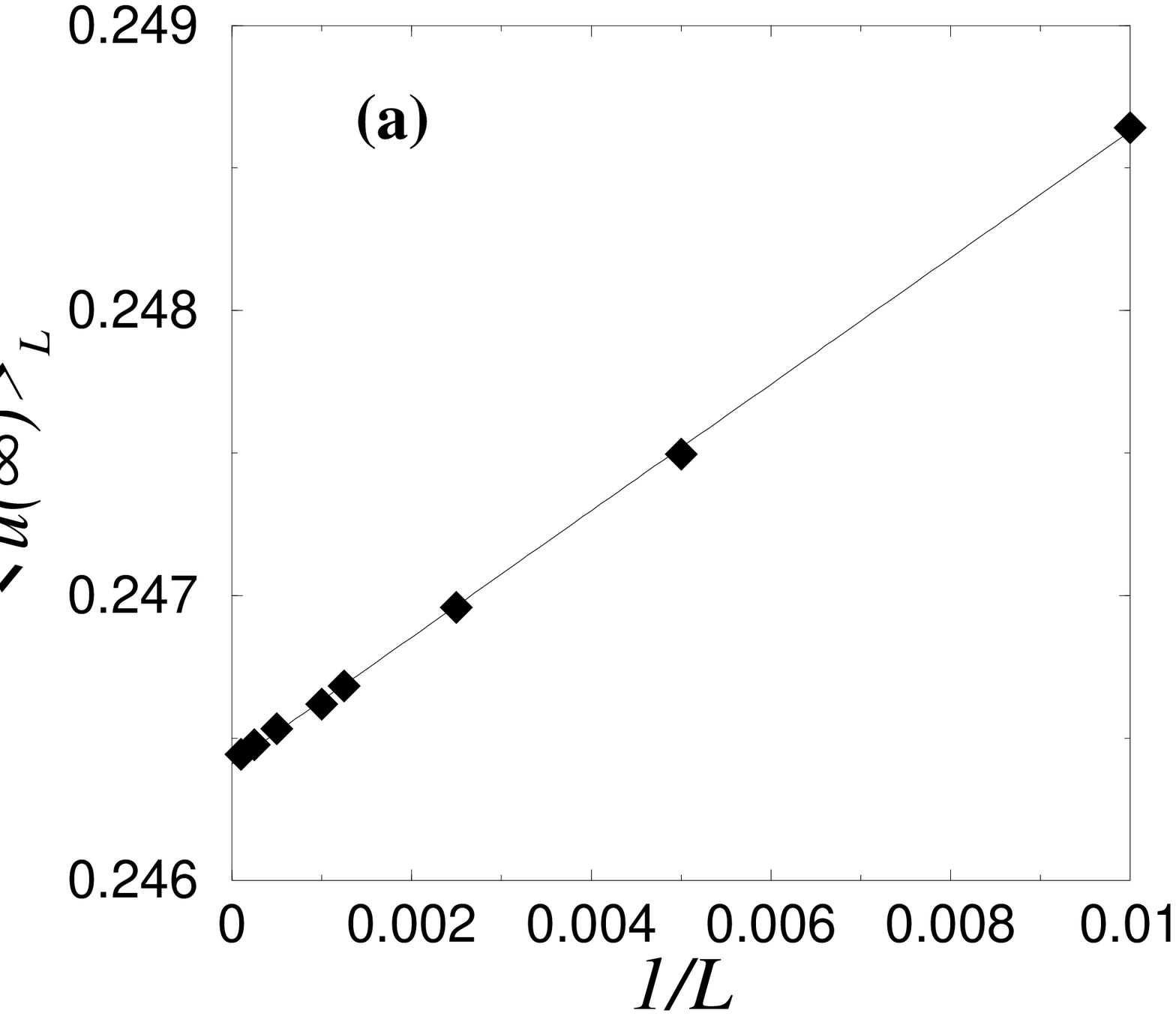}
\includegraphics{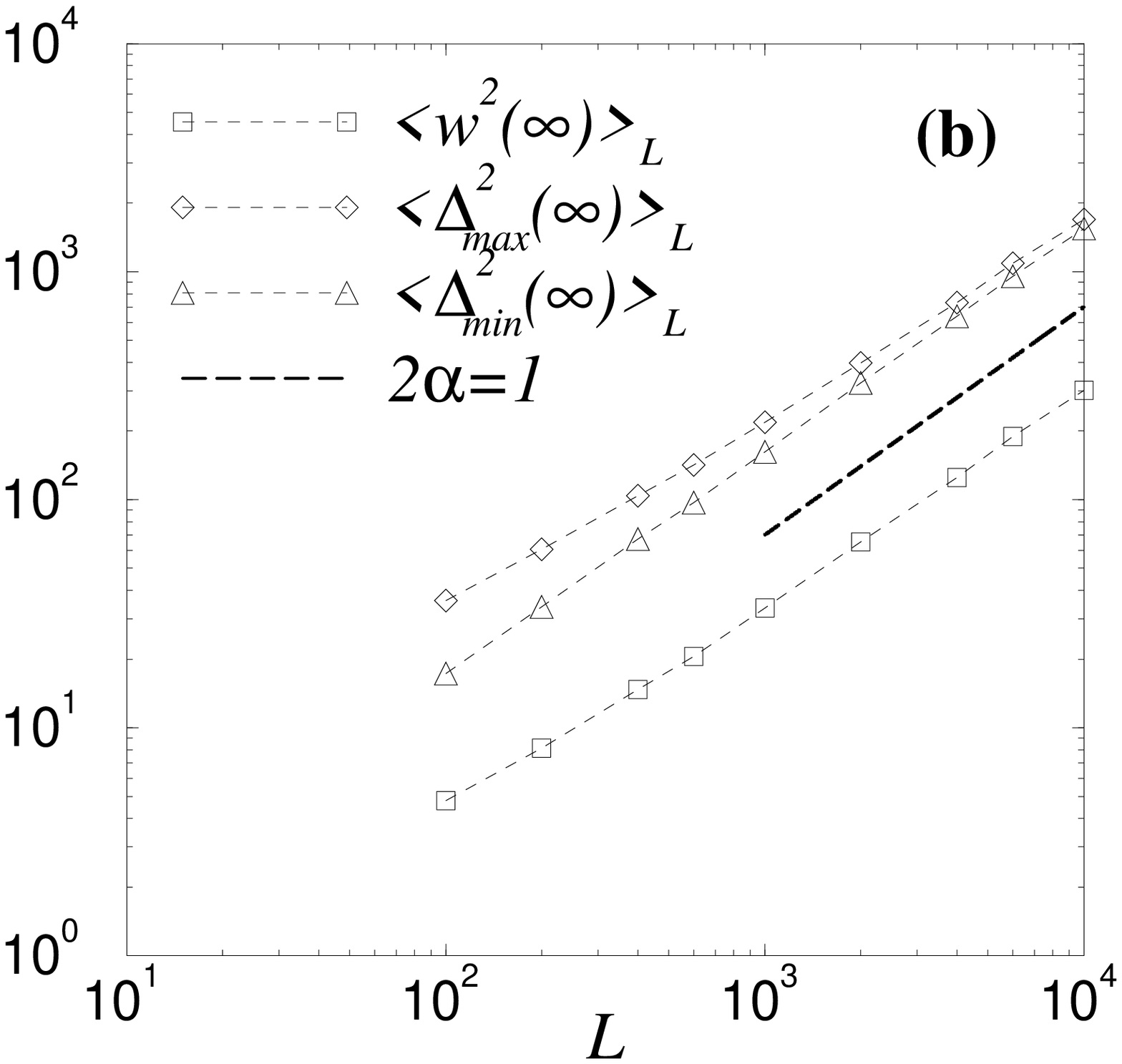}
\vspace*{3.0truecm}
\caption{\small
(a) Steady-state utilization (average rate of progress) as function
of the inverse system size.
(b) Steady-state average width and extremal fluctuations of the time
horizon. Note the log-log scales. The dashed line corresponds to a power law
with the exact KPZ exponent $2\alpha$$=$$1$. Both graphs for the
one-dimensional one site per PE basic conservative scheme.}
\label{fig5}
\end{figure}

Based on our results that the simulated time horizon exhibits
KPZ-like kinetic roughening, we now address the implications for the
scalability of conservative PDES schemes.
We already argued that KPZ universal surfaces evolve toward a steady
state where the coarse-grained local slopes,
$\partial\hat{\tau}/\partial x$,
become independent [see Eq.~(\ref{EW}) in particular].
This ensures that
there is a finite density of local minima, so that the PDES algorithm
progresses
at a non-vanishing rate in the limit of an infinite number of PEs.
Just as important for the utilization are the finite-size effects.
Using exact calculations based on the
Edwards-Wilkinson Hamiltonian in one dimension \cite{TORO00}
and scaling arguments in higher dimensions \cite{KRUG90}, one can obtain the
universal finite-size effects for the growth rate of generic KPZ-like
processes
\begin{equation}
\langle u(\infty)\rangle_L \simeq \langle
u(\infty)\rangle_{\infty} + \frac{\mbox{const.}}{L^{2(1-\alpha)}} \;.
\label{util_scale}
\end{equation}
The above equation can be used to estimate the utilization (average
rate of progress of the simulation) and to extrapolate to the
value for the infinite
number of PEs, $\langle u(\infty)\rangle_{\infty}$.
Equation~(\ref{util_scale}) for the finite-size effects is in full
agreement with the simulations [Fig.~\ref{fig5}(a)] and yields
$\langle u(\infty)\rangle_{\infty}$$=$$0.246461(7)$ for the one
dimensional case ($\alpha$$=$$1/2$).
While the actual asymptotic value of the density of
local minima depends on ``microscopic'' measures, whether its
asymptotic value vanishes or not, is fully governed by macroscopic
characteristics and the corresponding universality class
\cite{KORN00_PRL,TORO00}. We again emphasize that for
the KPZ class this asymptotic value is {\em non-zero}.

The above findings for the density of local minima, which
determines the average rate of progress of the simulation, imply
that the ``simulation part'' of the conservative scheme is
scalable. That is, if we run the simulation for long times, the
average progress rate approaches a constant. However, the kinetic
roughening exhibited by the time horizon has a disturbing implication:
in the steady state the
width (spread) of the simulated time horizon {\em diverges} with
the number of PEs as
\begin{equation}
\langle w^2(\infty) \rangle_L \sim L^{2\alpha} \;.
\end{equation}
This scaling behavior for large $L$ is also confirmed by
simulations [Fig.~\ref{fig5}(b)], and it is contrary to the conclusions of
Ref.~\cite{GREEN96}. This property creates an additional difficulty
for collecting statistics (e.g., to perform simple averages) ``on
the fly'' during the course of the simulation. The diverging
width means that the memory requirement {\em per PE}, for
temporarily storing (buffering) data, diverges as we increase the number of
PEs. In this sense we may call the ``measurement part'' of the
bare conservative scheme asymptotically nonscalable. Thus, in an
actual application, the programmer must implement some global
synchronization or a moving ``window" with respect to the global
minimum of the time horizon \cite{KORN02_ACM}.  However such a 
``window'' can have negative effects on the large $L$ utilization.  

Along these lines of questioning, we are also interested in the
{\em extremal} fluctuations of the time horizon. Namely, what is
the typical size of the largest fluctuations above and below the
mean,
$\Delta_{\rm max}(t)$$\equiv$$(\tau_{\rm max}(t) - \bar{\tau}(t))$
and $\Delta_{\rm min}(t)$$\equiv$$(\bar{\tau}(t) - \tau_{\rm min}(t))$,
where $\tau_{\rm max}(t)$ and $\tau_{\rm min}(t)$
are the {\em global} maximum and minimum simulated times among
$L$ PEs, respectively. We found that in the steady state
\begin{equation}
\left\langle\Delta^2_{\rm max}(\infty)\right\rangle_L \sim
\left\langle\Delta^2_{\rm min}(\infty)\right\rangle_L \sim L^{2\alpha} \;,
\label{extremal}
\end{equation}
i.e., they scale the {\em same way} as the width $\langle
w^2(\infty)\rangle_L$
[Fig.~\ref{fig5}(b)]. This should not come as a surprise, since this
surface is highly correlated, dominated by long-wavelength
fluctuations spreading over macroscopic length scales. This
finding, again, is consistent with the extremal fluctuations found
for general KPZ surfaces \cite{SHAPIR01}.

The above universal characteristics hold for the sensible and more
efficient many sites per PE case and/or when the interaction and
the corresponding communication pattern extends beyond
nearest-neighbor PEs (but is still short ranged with a finite
cutoff). For the case of many site per PE, however, the saturation
occurs at a later time and the time horizon exhibits a
substantially larger width \cite{KOLA_un}.

\vspace{0.4truecm}
\noindent {\bf SUMMARY AND OUTLOOK}

\vspace{0.3truecm}
We have studied the statistical properties of the basic conservative
parallel scheme for regular lattice topologies. We found that the
evolution of the simulated time horizon belongs to the well-known
KPZ dynamic universality class of non-equilibrium surfaces. This type of
growth is characterized by a {\em non-zero} density of local
minima, implying a non-zero rate of propagation in the
limit of an infinite number of PEs.
We also determined the asymptotic finite-size
corrections to this constant when the number of PEs is large but finite.
Thus, the ``simulation'' part of the algorithm is scalable.
Further, we showed that the spread (width) of the time horizon
approaches a finite constant for a {\em finite} number of PEs, but
this constant {\em diverges} as a power law in the limit of an
infinite number of PEs. The same holds for the extremal fluctuations
above and below the mean. This ``macroscopic''
roughness of the simulated time horizon means that the ``measurement part"
of the bare conservative scheme is not scalable. That is, there
is an extra difficulty associated with taking statistical measurements ``on
the fly''. Intermittent data on each PE haved to be stored until all
PEs reach the simulated time instant at which some statistics
collection (e.g., simple averaging over the full physical
application) is to be performed. The diverging spread of the time
horizon implies a diverging storage need for this purpose
on every PE. Thus, the programmer has to implement some global
synchronization or windowing technique to limit the spread of the
simulated time horizon in order not to exceed the memory
constraint. By knowing exactly the finite-size dependence of the
spread, for fixed $L$ one can determine the optimal time between global
synchronizations or the optimal window size.

Our findings are universal in the sense that they hold for any
{\em short-range} ``interaction" topology for PEs on regular
lattices. They are also valid in the case when each PE carries a
block of sites. The {\em asymptotic} scaling behavior is again
governed by the KPZ exponents, in such a way that for larger and
larger blocks, there is a crossover from the almost ``random
deposition'' \cite{BARA95} to KPZ-like growth at a later and later
time.

We must mention that there have been earlier attempts to theoretically
describe the scalability of the basic conservative PDES scheme.
To obtain an analytically tractable scalability model, Greenberg
et.al \cite{GREEN96} introduced the $K$-random model. In this model
at each update attempt, PEs compare their local simulated times to the
local simulated times of $K$ {\em randomly} chosen PEs (rechosen
at every update attempt). They showed that in the
$t$$\to$$\infty$, $N$$\to$$\infty$ limit the average rate of
progress of the simulation converges to a {\em non-zero} constant,
$1/(K+1)$. They also showed that the evolution of the
time horizon converges to a traveling-wave solution described by a
{\em finite width} of the distribution of the local times.
Finally, they suggested that convergence to a traveling-wave solution in
the $t$$\to$$\infty$, $N$$\to$$\infty$ limit is universal and
applicable for regular lattices as well. In obtaining this conclusion
they made the assumption that replacing the ``interaction" between
nearest-neighbor PEs on a regular grid with the same interaction
between {\em randomly} chosen PEs does not change the universality
class of the time horizon. It does. Comparing the local time for
each PE to $K$ randomly chosen others essentially turns the model
in to a mean-field-like one where the time surface is short-range
correlated and has a finite width in the limit of an infinite number of PEs.
As we have shown in this paper, the time horizon of the conservative
PDES for regular lattices and short-ranged interactions with finite
cutoffs exhibits KPZ-like kinetic roughening..
This shows that the underlying communication topology of the PEs
has crucial effects on the ``universal'' characteristics of the simulated
time horizon.

Realizing the importance of the
communication topology of the PEs, we are currently investigating how to
turn the original conservative scheme on {\em regular lattices} into a fully
scalable one, where both the ``simulation" and the ``measurement''
parts are scalable, without the need for any global
synchronization or windowing technique. Results of these studies will be
published elsewhere.

\vspace{0.4truecm}
\noindent {\bf ACKNOWLEDGMENTS}

\vspace{0.3truecm}
We thank B.D.\ Lubachevsky and Z.~R\'acz for
useful discussions. We acknowledge the support of NSF through Grant
Nos. DMR-9871455, DMR-0113049, and the support of the Research
Corporation, Grant No. RI0761. Z.T. was supported by the
U.S. Department of Energy under Contract No. W-7405-ENG-36.

\end{document}